\documentstyle[11pt,paspconf,epsf]{article}

\begin{document}

\title{The Core of the Great Attractor; Is it behind the Southern Milky Way?}

\author{P.A. Woudt, A.P. Fairall}
\affil{Department of Astronomy, University of Cape Town,
    Rondebosch 7700, South Africa}

\author{R.C. Kraan-Korteweg}
\affil{Observatoire de Paris-Meudon, D.A.E.C., 5 Place Jules Janssen,
    92195 Meudon Cedex, France}

\begin{abstract}

\noindent
The nature and the extent of the Great Attractor has been the subject of much
debate, not in the least due to the unfortunate position of its central
part being behind the Milky Way.
We here present the latest results from our deep optical galaxy search in the
southern Milky Way. A full view of the southern hemisphere is emerging, 
revealing ACO 3627 as the most prominent concentration of galaxies in the 
southern sky. Our follow-up spectroscopic observations support the idea that 
ACO 3627 is the dominant component of a ``great wall''--structure, similar
to Coma in the (northern) Great Wall. 

\end{abstract}


\keywords{galaxy clusters: ICM, individual: ACO 3627, large scale structure}

\section{Introduction}

Dust and stars in the plane of the Milky Way create a Zone of Avoidance (ZOA) 
in the extra-galactic sky. As such our Galaxy is a natural barrier severely 
constraining studies of the large scale structures in the Universe, the 
peculiar motion of the Local Group and other streaming motions which are 
important for understanding formation processes in the Early Universe and
for cosmological models.

\noindent
In the southern Milky Way, the Great Attractor (GA) area is a region of 
special interest in this respect. Apparent as a large-scale systematic 
flow of galaxies towards $(\ell,b,v) \sim (320^{\circ}, 0^{\circ}, 
4500$ km s$^{-1})$ (Kolatt et al.~1995), this predicted mass excess 
is largely obscured from our view. 
Though it is generally believed to be an extended region ($\sim 40^{\circ} 
\times 40^{\circ}$) of moderately enhanced density (Lynden-Bell 1991, 
Hudson 1994 or Kolatt et al.~1995, cf. their Fig.~3b), no counterpart for the
centre of its gravitational potential could be identified.

\noindent
In an effort to reduce the size of the ZOA, we have been engaged for some
years now in a deep optical galaxy search in the southern Milky
Way: $265\deg \la \ell \la 340\deg, |b| \la 10\deg$ (see Kraan-Korteweg 1989 
and Kraan-Korteweg \& Woudt 1994a for details on the galaxy search). 
This recently led to the recognition that the cluster ACO 3627 is
a very massive, nearby cluster of galaxies at the core of the GA 
$(\ell,b,v) = (325^{\circ}, -7^{\circ}, 4882$ km s$^{-1})$ (Kraan-Korteweg et 
al.~1996a), emphasizing therewith that gravitationally influential
constituents of the Local Universe are indeed hidden by the Milky Way -- their
appearance inconspicuous due to the obscuring veil of the Galaxy.

\noindent
We here present an ``all-sky distribution'' of the optically detected galaxies
in the southern hemisphere, and a discussion of ACO 3627 and its
role within the known extra-galactic structures.

\section{Partially obscured galaxies behind the Southern Milky Way}

One of the prime objectives of our deep optical galaxy search behind the
southern Milky Way, or other galaxy surveys in the ZOA,
is the derivation of a uniform whole-sky distribution of galaxies, i.e. as
would have been seen in the absence of the obscuring layers of the Milky Way
(Lynden-Bell 1994).

\begin{figure}[h]
\hfil \epsfxsize 11.3cm \epsfbox{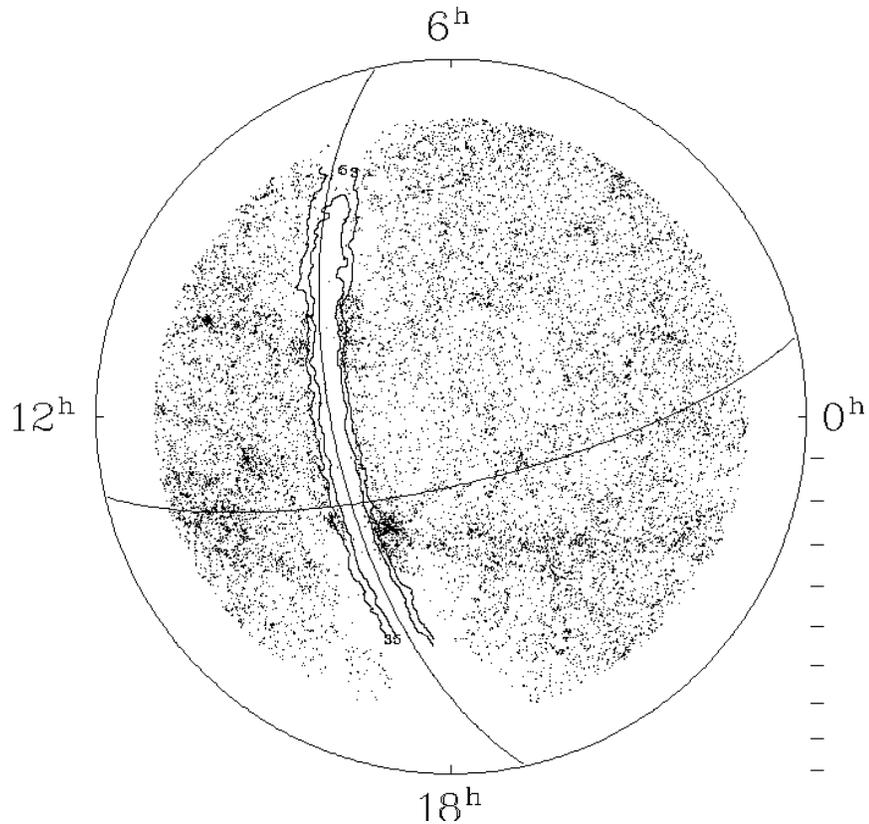}
\hfil
\caption
{An equal area distribution of Lauberts galaxies ($D \ge 1$\arcmin)
in the southern sky combined with galaxies from our survey that, when corrected
for the galactic foreground extinction, are subjected to the same selection 
criterion. Both the Galactic Plane (N-S) and the Supergalactic Plane (E-W)
are drawn. The contours are lines of equal extinction, deduced from the
HI column densities. The outer contour marks our completeness limit (see text).
The tick marks show various declinations, from 0$^{\circ}$ (outer circle) to 
-90$^{\circ}$ (centre) in steps of 10$^{\circ}$. }
\end{figure}

\noindent
As a first step in obtaining an all-sky distribution of galaxies in the 
southern sky, we have corrected the diameters of our newly found galaxies
for the diminishing effects of the extinction
using Cameron's (1990) law, the HI-column
densities (Kerr et al.~1986) as a tracer of the dust distribution and
the formalism of Burstein \& Heiles (1982), assuming a constant gas-to-dust 
ratio. From the 10276 galaxies found to date with $D \ge 0$\farcm$2$, 
1949 galaxies would have appeared in the 
Lauberts catalogue (Lauberts 1982, $D_{0} \ge 1$\farcm$0$) were it not for
the Milky Way. 
Of those 1949 galaxies, only 258 had been previously catalogued by Lauberts. 
The combined distribution of all the galaxies with $D_{0} \ge 1$\farcm$0$
is shown in fig.~1.

\noindent
The contours drawn in figure 1 are deduced from the HI-column densities. 
The outer contour ($A_B = 3$\fm$2$) corresponds to an apparent diameter 
reduction for spiral galaxies of $f=5$ (Cameron 1990) and demarcates our 
completeness limit: below this line a galaxy with 
$D_{0} = 1$\farcm$0$ would be smaller than our diameter limit and 
therefore go unnoticed.
Above this band distinct concentrations and filaments can be recognized.
The inner contour ($A_B = 5$\fm$0$) shows the remaining ZOA. Below this contour
the Milky Way remains opaque.

\noindent
The most prominent concentration of galaxies in the southern sky is centred 
on ACO 3627 $(\alpha, \delta) = (16^h10^m, -60^{\circ}48')$ (Abell et 
al.~1989), just below the Galactic Plane close to the cross-section with 
the Supergalactic Plane. Other major concentrations of galaxies in figure 1
are the Hydra cluster $(10^h35^m, -27^{\circ}16')$ and the Centaurus 
cluster $(12^h46^m, -41^{\circ}02')$.

\section{ACO 3627}

The cluster properties of ACO 3627 proof this cluster to be a rich,
massive cluster of galaxies comparable to the well known Coma cluster,
but closer in redshift-space ($4882$ km s$^{-1}$ vs. $6960$ km s$^{-1}$) 
(Kraan-Korteweg et al.~1996a). This view is supported by the
recent observations of ACO 3627 with the ROSAT PSPC, which finds this cluster
to be the 6$^{th}$ brightest X-ray cluster in the ROSAT All Sky Survey
(B\"ohringer et al.~1996).

\noindent 
In figure 2 we have superimposed the X-ray contours
on the central part of ACO 3627, as reproduced from the ESO/SRC IIIaJ 
film copy of field 136. The contours of the 843 MHz radio continuum emission
from 2 galaxies in ACO 3627 (Jones \& McAdam 1992) are also plotted. 
Both the X-ray and the radio continuum emission provides substantial 
additional information on the cluster morphology and the Intra-Cluster
Medium (ICM) of ACO 3627.

\noindent
Like many rich clusters, ACO 3627 shows distinct subclustering
in the X-ray morphology. In figure 2, the X-ray shows extended emission towards
the SE corner (bottom-left). After the subtraction of a spherical symmetric
model, a residual component remains, suggestive of a subcluster in the 
process of merging (B\"ohringer et al.~1996, Kraan-Korteweg 
et al.~1996b). This view is bolstered by the radio-morphology of PKS1610-608. 
This strong (one of the 20 strongest extra-galactic radio sources) 
wide-angle-tail source embraces the merging system (cf. figure 3 of 
Kraan-Korteweg et al.~1996b), revealing a strong motion of the cluster gas 
in a subcluster merger (Jones \& McAdam 1996 and Burns et al.~1994).

\noindent
ACO 3627 harbours two other interesting galaxies: the head-tail radio-source
B1610-605 and a Seyfert 1.3 galaxy. The head-tail source shows an alignment
with the 3rd contour of the main X-ray component over nearly its full extent.
Jones \& McAdam (1996) find a particularly high pressure near the peak of
B1610-605, indicative of  the ram pressure as the galaxy moves through 
the ICM. The strong X-ray point source in the top-right corner of figure 2 is
a Seyfert 1.3 galaxy, first discovered by spectroscopy (Woudt et al.~1996).

\medskip

\begin{figure}[h]
\hfil
\caption
{The central part (56\arcmin x 56\arcmin) of the cluster ACO 3627 as 
reproduced from field 136 of the IIIaJ copy of the ESO/SRC survey. 
Superimposed are the X-ray contours (thick contours) from the ROSAT 
PSPC observations (B\"ohringer et al.~1996), 
and the 843 MHz radio continuum emission of the wide-angle-tail
radio galaxy PKS1610-60.8 and the head-tail radio-source B1610-60.5 (Jones
\& McAdam 1992).
The strong X-ray point source in the top-right corner is a Seyfert~1 
galaxy, also a member of ACO 3627.}
\end{figure}

\vfill \eject

\section{ACO 3627 and the Great Attractor}

Redshift observations are required to map the newly found galaxies in
3 dimensions. For this we have used three complimentary approaches:

\begin{itemize}
\item{Multifiber spectroscopy with OPTOPUS and MEFOS at the 3.6-m telescope
of ESO for galaxies in the densest regions (Cayatte et al.~1994, Kraan-Korteweg
et al.~1994 and Felenbok et al.~1996)}
\item{Individual spectroscopy of the brightest galaxies with the 1.9-m 
telescope of SAAO (Kraan-Korteweg et al.~1995)}
\item{21-cm Observations of extended LSB spirals with the 64-m Parkes
radio telescope (Kraan-Korteweg et al.~1996b)}
\end{itemize}

\begin{figure}[h]
\hfil \epsfxsize 11.3cm \epsfbox{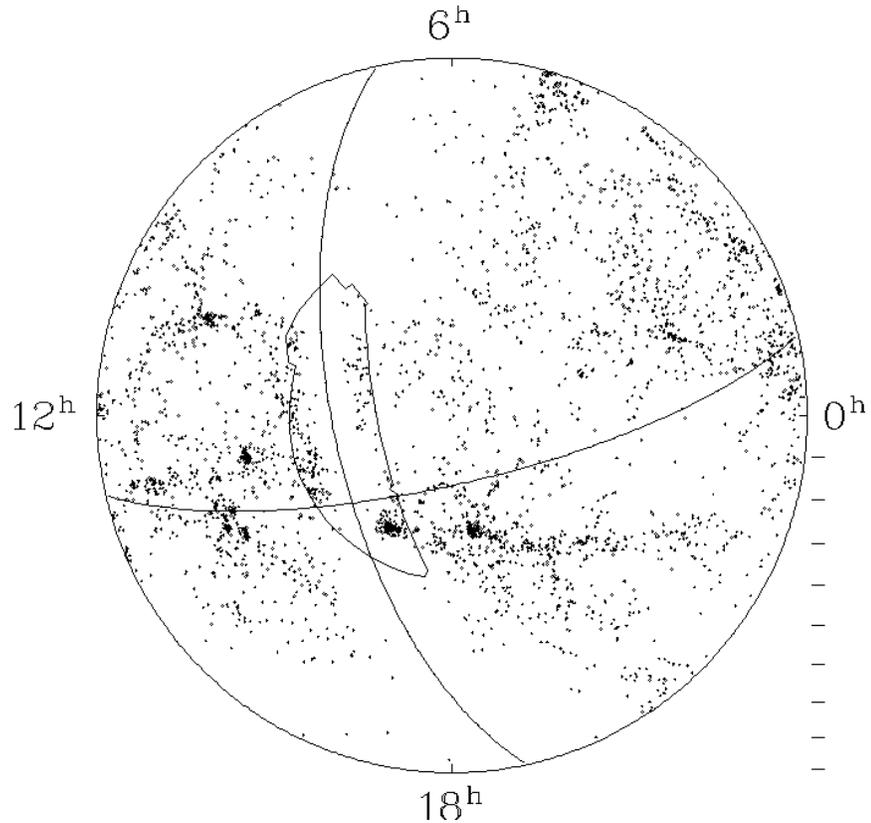}
\hfil
\caption
{An equal area distribution of all the galaxies in the southern sky 
within redshift interval: $3500 \le v_{obs} \le 6500$ km s$^{-1}$.
Published redshifts are taken from the Southern Redshift Catalogue
(Fairall 1996) and
are complemented by data from our follow-up spectroscopic survey of 
galaxies behind the southern Milky Way. Our search area is outlined, the 
Galactic and Supergalactic Plane are marked. The tick marks on the right
are as in fig.~1.}
\end{figure}

\noindent
Figure 3 shows an equal area distribution of all the galaxies in the 
southern hemisphere with $3500 \le v_{obs} \le 6500$ km s$^{-1}$,
i.e.~centred on the approximate redshift-distance of the GA. 

\noindent
In this redshift 
range ACO 3627 is seen to be the central, dominant component of a ``great 
wall''-like structure, a broad structure including the Indus group $(\alpha, 
\delta, v) = (21^h03^m, -47^{\circ}21', 4842$ km s$^{-1})$, the Pavo 
cluster $(18^h43^m,$ $-63^{\circ}23',$ $4167$ km s$^{-1})$ crossing the 
central part of the GA region towards $(12^h$, $-55^{\circ}$, 
$5500$ km s$^{-1})$ where it possibly merges
with the Vela supercluster at $(10^h15^m,$ $-49^{\circ}20'$, 
$6000$ km s$^{-1})$ (Kraan-Korteweg \& Woudt 1994b).

\noindent
The here emerging picture of the galaxy distribution supports the earlier
suggestions that the GA is an extended region of moderately enhanced density
with the cluster ACO 3627 now marking the previously unidentified core of
the Great Attractor.

\section{Future prospects}

\noindent
Due to its proximity, the cluster ACO 3627 provides an excellent sample
to investigate environmental effects on its galaxies. Our presently ongoing
ATCA observations of four selected regions in ACO 3627, in combination
with our other observations, will allow a detailed study of the ICM in
ACO 3627.

\noindent
Is ACO 3627 at rest with respect to the Cosmic Microwave Background? Here,
the Tully-Fisher relation for spirals and the $D_{n} - \sigma$ analysis of 
early type galaxies -- although difficult due to the foreground extinction
(Mould et al.~1991) -- will give us the answer.

\noindent
A further reduction of the remaining ZOA ($|b| \le 5\deg$) will be achieved
with the forthcoming MultiBeam ZOA-survey at the Parkes 64-m radio
telescope. It will yet be another step towards a true all-sky distribution
of galaxies. Only then can we trace the full extent of the Norma
(ACO 3627) Supercluster.\\

\acknowledgments

PAW and APF are supported by the South African FRD. The research of RCKK is 
being supported with an EC-grant. This research has made use of the NASA/IPAC
Extragalactic Database (NED), which is operated by the Jet Propulsion 
Laboratory, Caltech, under contract with the National Aeronautics and Space
Administration.

\end{document}